\documentclass[amsmath,amssymb,floatfix,groupedaddress,twocolumn,prl]{revtex4}

\usepackage{graphicx}

\newcommand{\msub}[1]{_{\mathrm{#1}}}

\begin{document}

\title{An optical fiber-taper probe for wafer-scale microphotonic device characterization}

\author{C.~P.~Michael}
\email{cmichael@caltech.edu}
\affiliation{Department of Applied Physics, California Institute of Technology, Pasadena, CA 91125}
\author{M.~Borselli}
\affiliation{Department of Applied Physics, California Institute of Technology, Pasadena, CA 91125}
\author{T.~J.~Johnson}
\affiliation{Department of Applied Physics, California Institute of Technology, Pasadena, CA 91125}
\author{C.~Chrystal}
\affiliation{Department of Applied Physics, California Institute of Technology, Pasadena, CA 91125}
\author{O.~Painter}
\affiliation{Department of Applied Physics, California Institute of Technology, Pasadena, CA 91125}

\begin{abstract} 
A small depression is created in a straight optical fiber taper to form a local probe suitable for studying closely spaced, planar microphotonic devices.  The tension of the ``dimpled'' taper controls the probe-sample interaction length and the level of noise present during coupling measurements.  Practical demonstrations with high-$Q$ silicon microcavities include testing a dense array of undercut microdisks (maximum $Q = 3.3${}$\times$10$^6$) and a planar microring ($Q = 4.8${}$\times$10$^6$).
\end{abstract}
\date{\today}

\maketitle

%
%
%

\section{Introduction}
In microelectronics manufacturing, nondestructive parametric testing using metal probe tips greatly increases fabrication yield through statistical process control~\cite{Montgomery_quality_control}.  For testing of glass and semiconductor photonic lightwave circuits (PLCs), many methods exist for the coupling of light into and out of on-chip waveguides~\cite{Pavesi-Si_photonics_2004}. However, no simple, local probe exists for wafer-scale, nondestructive, optical characterization of on-chip components.  Traditional optical coupling methods include end-fire or butt coupling~\cite{ol28-1302,ieee-ofc2003-v1-p249,oe11-3555} and prism-based coupling~\cite{optcommun113-133,josaA17-802}.  End-fire coupling from free-space or optical fibers can be made highly efficient, even to high-index contrast semiconductor waveguides, through the use of tapered waveguide sections~\cite{ol28-1302,ieee-ofc2003-v1-p249,oe11-3555} or other non-adiabatic mode converters~\cite{ieee-jlt16-1228,ieee-jlt16-1680}, but they are limited to coupling at the periphery of the chip where a cleaved facet can be formed.  Evanescent-coupling methods involving conventional prism couplers, angled-fiber tip couplers~\cite{ol24-723}, eroded-fiber couplers~\cite{ol20-813}, and optical fiber tapers~\cite{ol22-1129,ieee-ptl11-686,ol25-260}, can provide effective coupling to and from on-chip waveguides, but these probes are less suited to wafer-scale coupling to micron-scale photonic elements due to their macroscopic extent in one or both in-plane dimensions.  Evanescent coupling techniques also rely on phase-matching to obtain highly efficient coupling~\cite{prl91-043902,oe13-801,prb72-205318}, which can be difficult (although not impossible~\cite{josab20-2274,oe13-801}) to satisfy for semiconductor-based microphotonic chips.  Other methods of coupling light onto photonic chips for characterization purposes involve dedicated on-chip testing structures such as in-plane grating couplers~\cite{paddon_input_coupler_patent}.  These couplers typically also involve specialized processing to achieve high coupling efficiency:  blazed gratings~\cite{apl77-4214}, a combination of lateral and vertical Bragg reflectors~\cite{ieee-jqe38-949}, or additional overlayers~\cite{oe14-11622}.     

We present a variant of the silica optical fiber taper evanescent-coupler that is designed for rapid, wafer-scale diagnostic testing of on-chip photonic components such as waveguides and resonant filters.  Previous work involving straight fiber tapers required devices to be elevated by several microns above the chip surface to prevent parasitic coupling to the surrounding substrate.  Macroscopically curved fiber taper probes~\cite{apl87-131107,ieee-jqe42-131,oe14-1070,oe15-1267} have been demonstrated which reduce parasitic loss into the substrate.  However, they tend to be less mechanically stable than their tensioned straight-taper counterparts and suffer from noise induced by fluctuations in the taper's position.  In this work we have developed a microscopic ``dimpled'' fiber taper probe which allows for low-noise local probing of individual devices on a wafer.  By increasing the tension in the taper, fluctuations in the taper-chip gap can be greatly reduced to the levels present in straight fiber taper measurements.  To demonstrate the utility of the dimpled taper optical probe, we describe the characterization of two types of devices on a silicon-on-insulator (SOI) wafer platform: a dense two-dimensional array of high-$Q$ silicon microdisk resonators and, secondly, a planar microring resonator.

\section{The Dimpled Fiber-Taper Probe}

\begin{figure}
\begin{center}
\includegraphics[width=\columnwidth]{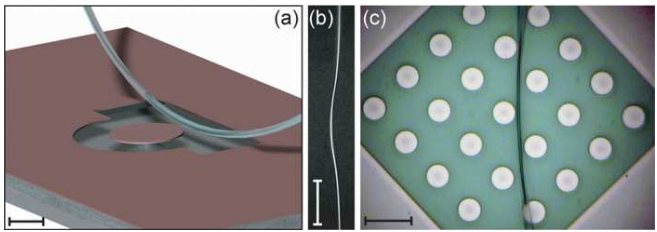}
\caption{(a) Schematic of a ``dimpled'' taper coupled to an undercut microdisk.  (b) Optical image of the taper probe.  The taper diameter at the center of the dimple is $\sim$1.2\,$\mu$m.  (c) At the center of a 5$\times$5 array, the dimpled taper probe is critically coupled to a microdisk but not coupled to any of the neighboring disks.  The scale bars are (a) 5\,$\mu$m, (b) 50\,$\mu$m, and (c) 20\,$\mu$m.}
\label{dimple}
\end{center}
\end{figure}

The dimpled fiber taper probe is made from a standard straight fiber taper that is pressed against a mold and heated.  We form ``straight'' fiber tapers by simultaneously heating and pulling standard telecommunication fiber (specifically SMF-28e).  By slowly thinning the fiber, the fundamental core-guided fiber mode is adiabatically converted to the fundamental taper mode with evanescent tails that extend significantly into the surrounding medium.  After mounting the taper in a U-bracket~\cite{ol29-697}, the narrowest part of the taper is pressed against a silica mold with the desired radius of curvature; a bare optical fiber with a radius of approximately 62\,$\mu$m is used as the mold in these experiments.  The taper and mold are heated with a hydrogen torch and allowed to cool.  After detaching the fiber from the mold, the taper retains an impression of the mold, Fig~\ref{dimple}(b), which forms a global minimum with respect the rest of the taper.  The dimpling process introduces negligible additional loss, and the total loss of the dimpled taper is typically less than 0.5\,dB relative to the un-pulled optical fiber. 
Using a specially designed U-mount with a set screw to control the tensioning, varying the taper's tension changes the radius of curvature of the dimple.  Under high tension, the dimple becomes very shallow but never completely straightens.  After dimpling, the probe is mounted onto a three-axis 50-nm-encoded stage and is fusion-spliced into a versatile fiber-optic setup.  During testing, devices are placed in the near-field of the probe, as in Fig.~\ref{dimple}(a,c); adjustments to a pair of goniometers ensure the straight run of the taper is parallel to the sample surface. 

\begin{figure}
\begin{center}
\includegraphics[width=\columnwidth]{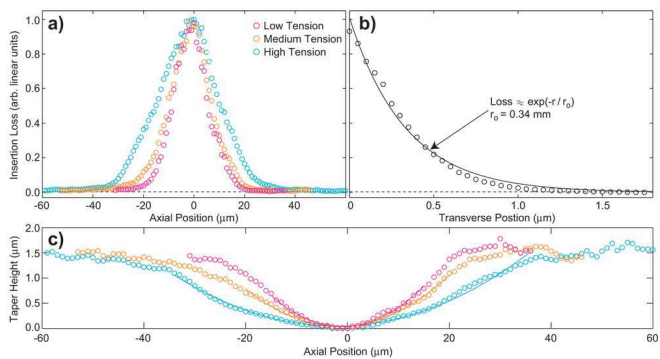}
\caption{Non-resonant insertion loss (a) as a function of axial position as a narrow cantilever is moved along the taper length and (b) as a function of transverse position as the dimple is raised above a mesa.  (c) Inferred dimple taper profile at ``low,'' ``medium,'' and ``high'' tension.}
\label{profile}
\end{center}
\end{figure}

Measurement of the non-resonant insertion loss as the waveguide is moved relative to nearby semiconductor microstructures gives the effective interaction length and profile of the local probe.  First, we record the loss as a 1.6-$\mu$m wide GaAs cantilever is scanned along the taper's length while holding the taper at a fixed height.  At tensions used in standard testing, Fig.~\ref{profile}(a) shows only $\sim$20\,$\mu$m (full width at half max) of the taper at the bottom of the dimple is close enough to interact with the sample.
Second, the loss is measured as a function of the probe's height above a 11.6-$\mu$m wide GaAs mesa.  By assuming an exponential vertical dependence for the insertion loss [Fig.~\ref{profile}(b)], we convert the loss's axial dependence [Fig.~\ref{profile}(a)] into the probe's ``near-field'' profile [Fig.~\ref{profile}(c)]---i.e. the height of the taper relative to the lowest point of the dimple.  Since only the lowest part of the dimple interacts with the sample, this method can only determine the taper's profile within $\sim$1.25\,$\mu$m of the surface.  Fitting the profiles determines the effective probe radius is 159\,$\mu$m, 228\,$\mu$m, and 498\,$\mu$m at low, medium, and high tension, respectively.  These radii differ from the mold radius ($\sim$62\,$\mu$m) due to tensioning of the taper and how the fiber detaches from the mold after heating.

To study the resonators in the following demonstrations, the devices were excited using fiber-coupled swept tunable-laser sources (spanning 1423--1496\,nm and 1495--1565\,nm, linewidth $< 300$\,kHz over the 25-ms time scale needed to scan across a high-$Q$ resonance) and a paddle-wheel polarization controller to selectively couple to TE-like and TM-like modes.  To measure the intrinsic quality factor, the cavities are weakly loaded by the dimpled probe and excited at low power.  Without any optical amplification, the signal is acquired using a high-speed photodetector, electrically amplified using a low-noise analog preamplifier, and then is saved by a analog-to-digital converter.  For measured $Q > 10^6$ (linewidth $\delta\lambda \lesssim 1.5$\,pm), the linewidth measurement is immediately calibrated with a fiber-optic Mach-Zehnder interferometer to an accuracy of $\pm$0.01\,pm.

\section{Noise Measurements}

\begin{figure*}
\begin{center}
\includegraphics[width=1.8\columnwidth]{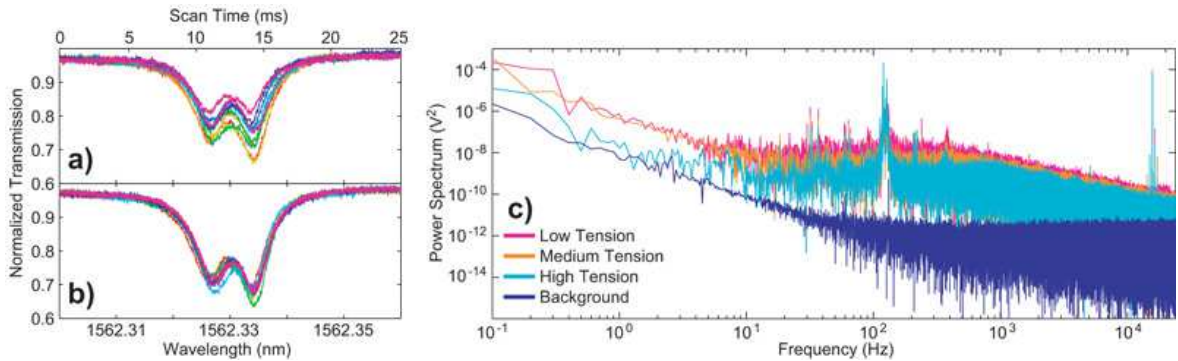}
\caption{Reducing noise through higher taper tension.  Without averaging multiple scans, ten consecutive traces of a microdisk ``doublet'' resonance~\cite{oe13-1515} display greater displacement noise at low tension (a) than at higher tension (b).   (c) Noise power spectra with the same tensions as in Fig.~\ref{profile}(a,c).}
\label{noise}
\end{center}
\end{figure*}

Because evanescent coupling to fiber tapers is exponentially dependent on position, fiber-taper measurements are very susceptible to any noise sources that produce physical displacements of the taper.  For straight tapers, increasing tension to reducing these fluctuations is common, and the U-mount~\cite{ol29-697} naturally provides the appropriate tautness.  Isolating the measurements from stray air currents is also imperative---typically all testing is conducted in a continuously N$\msub{2}$-purged enclosure.  Under standard testing conditions at low dimple-taper tension [Fig.~\ref{noise}(a)], coupling to the mode of a microdisk resonator [see Fig.~\ref{dimple}(a) and Section~\ref{sec:array}] varies significantly between consecutive scans.  Increasing the tension makes the coupling depth much more reproducible, as in Fig.~\ref{noise}(b).   At tensions that give acceptable noise levels, the depth of the dimple is still adequate for testing densely-spaced planar devices.

To quantitatively study the noise, we measure non-resonant insertion loss as a function of time.  The dimple is placed above the etched GaAs mesa so that approximately 60\,percent of the incident power is coupled into the substrate.  The mesa structure assures a constant 11.6-$\mu$m interaction length for different taper tensions.  We minimize the electrical noise contribution by maximizing the incident optical power in order to decrease the needed electrical gain and filtering.  We also eliminate extraneous noise sources (unused computers, monitors, overhead lights, etc.) and turn off the N$\msub{2}$ flow into the testing enclosure.  To obtain a background spectrum that is independent of any taper displacement, the dimple is raised so no power is coupled into the substrate, and then the power is attenuated to give the same output voltage from the detector.  The resulting noise power spectra in Fig.~\ref{noise}(c) reveal increasing tension reduces broadband noise between approximately 10 and 1000\,Hz, reflecting the relevant time-scales for scanning across a high-$Q$ resonance.   The series of high-frequency peaks at $\sim$15.8\,kHz occur at the pulse-position-modulation clock frequency of the stage motor controller.  The dominant spike at low frequencies is bimodal with peaks at $\sim$120\,Hz and $\sim$130\,Hz with a total bandwidth of $\sim$20\,Hz.   The motor controller also contributes to noise in this band, but it is not the dominant noise source.  We hypothesize that electrical noise actuates the motors and drives low-$Q$ vibrational modes of the fiber taper.   By measuring insertion loss as a function of the dimple-substrate gap and comparing it to noisy time-domain transmission traces, we estimate the upper bound on fluctuations in the taper height to be $7.9\pm1.4$\,nm, which is consistent with our earlier measurements with straight tapers.   


\section{Characterization:  Microdisk Array}
\label{sec:array}

To demonstrate the dimpled taper's ability to test closely spaced devices, we study a 5$\times$5 array of silicon microdisks [Fig.~\ref{dimple}(c)] with disk diameters of 10\,$\mu$m and periodicity of 20\,$\mu$m---corresponding to an areal density of 2.5$\times$10$^5$\,cm$^{-2}$.  Undercut microdisks were chosen over planar resonators to ease phase matching between the cavity and taper modes.  The microdisks were fabricated from silicon-on-insulator with a 217-nm device layer [$\langle100\rangle$ orientation, p-type, 14--20\,$\Omega\cdot$cm] and a 2-$\mu$m SiO$_2$ buried oxide layer (BOX).  The resonators were defined using electron-beam lithography, resist reflow, and reactive ion etching; then the disks were partially undercut by etching the buried oxide using dilute HF.  The silicon surfaces are temporarily hydrogen passivated using repeated Piranha/HF treatments.  Long-term passivation is achieved using a 3-nm dry thermal oxide cap grown in O$_2$ at 1000$^\circ$C followed by a 3-hour anneal in N$_2$ at 1000$^\circ$C and then a 1.5-hour slow cool down in N$_2$ from 1000$^\circ$C to 400$^\circ$C.  For details on the lithography, chemical passivation, and oxide passivation, see Refs.~\cite{oe13-1515}, \cite{apl88-131114}, and \cite{MB_surface_passivation}, respectively.  

\begin{figure*}
\begin{center}
\includegraphics[width=1.8\columnwidth]{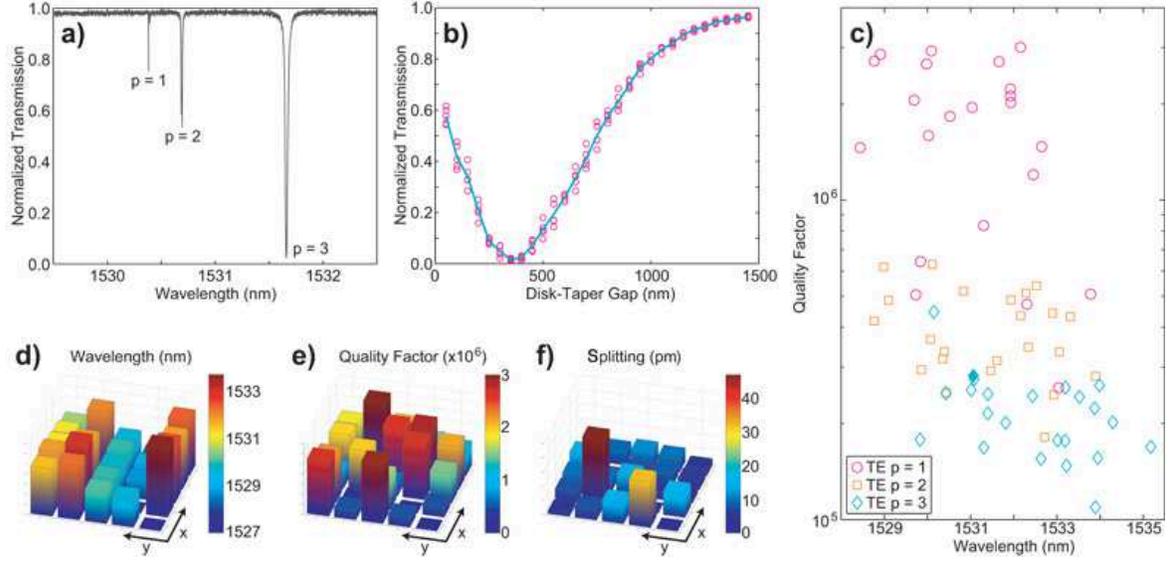}
\caption{(a) Sample transmission spectrum for a single microdisk.  (b) Coupling dependence on the disk-taper gap for a TE~$p=3$ mode of the device in Fig~\ref{dimple}(c).  (c) Distribution of wavelengths and quality factors for the TE~$p=1$--3 modes near 1532\,nm.  The solid diamond indicates the mode tested in (b).  Spatial distribution for the (d) wavelength, (e) quality factor, and (f) doublet splitting of the TE~$p=1$ modes.}
\label{array}
\end{center}
\end{figure*}

Near 1532\,nm, we track three TE-like modes of different radial orders [$p=1$--3 in Fig.~\ref{array}(a)] across all 25 disks in the array.    
One disk supported no high-$Q$ whispering-gallery modes in the range spanning 1495--1565\,nm, and we were unable to couple to the TE $p=1$ mode in two other disks---most likely because their $Q$ was too low to overcome the phase mismatch with the taper mode.   In Fig.~\ref{array}(b), varying the disk-taper coupling through their separation practically demonstrates the level of displacement noise present in these measurements; each circle represents the transmission minimum for an individual scan at the given probe position.  Table~\ref{array_table} summarizes the average measured wavelength ($\lambda\msub{o}$), quality factor, and doublet~\cite{oe13-1515} splitting ($\Delta\lambda$) for each mode; the distributions of wavelength and quality factor~\cite{footnote1} appear in Fig.~\ref{array}(a).
\begin{table*}
\begin{center}
\caption{Average mode parameters for microdisk array}
\label{array_table}
\begin{tabular}{c c c c c}
\hline\hline
Mode		& Observed		& $\lambda\msub{o}$ (nm) 	& Q						& $\Delta\lambda$ (pm)\\
\hline
TE $p=1$	& 22/25		& $1531.008\pm1.487$	    	&(1.73$\pm$0.93)$\times$10$^6$ 	& $11.31\pm10.12$  \\
TE $p=2$	& 24/25		& $1531.393\pm1.508$	    	&(3.95$\pm$1.32)$\times$10$^5$ 	& $10.93\pm5.60$  \\
TE $p=3$	& 24/25		& $1532.429\pm1.489$	    	&(2.19$\pm$0.70)$\times$10$^5$ 	& $10.70\pm5.77$  \\
\hline\hline
\end{tabular}
\end{center}
\end{table*}
The highest $Q$ for a single standing wave mode is 3.3$\times$10$^6$ with $Q/V = 2.3$$\times$10$^5$ for a calculated mode volume $V = 14.09$\,$(\lambda/n)^3$.   With minimal free-carrier absorption in the bulk~\cite{footnote2}
, the modal loss likely has significant contributions from both surface absorption and surface scattering since the ratio of the doublet splitting (related to the surface scattering rate) over the resonance linewidth varies from 3.1 to 28.1 for modes with $Q>10^6$.  The spatial arrangement of the mode parameters across the array [Fig.~\ref{array}(d--f)] shows a systematic change in $\lambda\msub{o}$ and more random variations in $Q$ and $\Delta\lambda$.  The $\lambda\msub{o}$ distribution implies the sample was slightly tilted with respect the beam writer's focal plane.  Similar geographic patterns exist for the parameters of the $p=2$ and $p=3$ modes.

\section{Characterization:  Planar Microring}

\begin{figure}
\begin{center}
\includegraphics[width=\columnwidth]{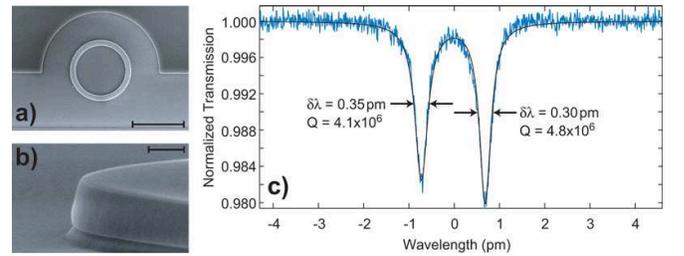}
\caption{SEM images of rings after the final chemical treatments and 30-nm thermal oxidation:  (a) top-view of a ring with a 20-$\mu$m diameter and 2-$\mu$m width and (b) side view showing smooth ring sidewalls and a slight BOX undercut due to the final chemical treatments.  The scale bars are (a) 20\,$\mu$m and (b) 200\,nm.  (c) Transmission spectrum of a high-$Q$ mode at $\lambda\msub{o} = 1428.7$\,nm in a ring with an 80-$\mu$m diameter and a 2-$\mu$m width.}
\label{ring}
\end{center}
\end{figure}

Testing planar devices is accomplished in the same fashion.  Non-undercut microring resonators, shown in Fig.~\ref{ring}(a), were fabricated from SOI with a 195-nm silicon device layer and a \mbox{3-$\mu$m} BOX.   
The same lithography, resist reflow, and dry etching procedure used for the microdisks~\cite{oe13-1515} was used to define the microrings although without the final HF undercut.  Repeated Piranha oxidations and HF dips are again used to chemically passivate the surfaces prior to thermal oxidation~\cite{apl88-131114}; these treatments also slightly undercut the resonators [Fig.~\ref{ring}(b)].  Finally, a 30-nm dry thermal oxide was grown as a capping layer, and the microring sample was annealed according to the same N$_2$ schedule as the microdisks~\cite{MB_surface_passivation}.  

Microrings are slightly more difficult to test with fiber tapers than undercut microdisks.  A large phase mismatch exists between the taper and microcavity because of the extra dielectric beneath the Si-core guided modes.  With the taper in contact with the ring, the coupling depth is more than sufficient to assess the devices' optical loss characteristics.  However, the coupling is not adequate to efficiently excite and collect emission from active devices~\cite{prb72-205318}.  For applications requiring high pump/collection efficiency, photonic crystal waveguides can be used to overcome the poor phase matching between the modes in the taper and the modes in the on-chip device~\cite{oe13-801}.

Figure~\ref{ring}(c) shows a transmission spectrum of a ring with an 80-$\mu$m diameter and 2-$\mu$m width after the final chemical treatments and thermal oxidation.  The measured quality factor of 4.8$\times$10$^6$ (loss coefficient $\alpha < 0.1$\,dB/cm) represents the highest quality factor for any planar microresonator to date.  Reproducing $Q$s found previously only in relatively thick and undercut silicon disks~\cite{oe13-1515} is promising for the future development of PLCs with high-$Q$ silicon microresonators integrated with bus waveguides.

\section{Conclusions}
Using a dimpled fiber taper waveguide, we have demonstrated a localized optical probe capable of testing dense arrays of planar devices.  Proper tensioning makes the dimpled taper more robust against fluctuations in position and decreases broadband noise.   Even without dedicated test structures to ease phase-matching constraints, the local dimpled-taper probe enables nondestructive wafer-scale optical characterization for manufacturer-level statistical process control.   Higher yields through low-cost testing will become increasingly important in a growing market where the burgeoning demand for bandwidth is making integrated micro-electronic-photonic solutions more attractive~\cite{inteltech8-129}.

%

\section*{Acknowledgments}

We thank M.~D.~Henry, K.~Srinivasan, and K.~Hennessy for fabrication assistance and M.~Hochberg and A.~Scherer for the SOI wafer used to fabricate the planar microring sample.  This work was supported by the DARPA EPIC program, contract number HR0011-04-1-0054.  For graduate fellowship support, we thank the Moore Foundation (CPM and MB), NSF (CPM), NPSC (MB), and HRL Laboratories (MB).


\begin{thebibliography}{10}
\newcommand{\enquote}[1]{``#1''}
\expandafter\ifx\csname url\endcsname\relax
  \def\url#1{\texttt{#1}}\fi
\expandafter\ifx\csname urlprefix\endcsname\relax\def\urlprefix{URL }\fi
\providecommand{\eprint}[2][]{\url{#2}}

\bibitem{Montgomery_quality_control}
D.~C. Montgomery, \emph{Introduction to Statistical Quality Control} (John
  Wiley \& Sons, Ltd., New York, 1991).

\bibitem{Pavesi-Si_photonics_2004}
L.~Pavesi and D.~J. Lockwood, eds., \emph{Silicon Photonics}, vol.~94 of
  \emph{Topics in Applied Physics} (Springer-Verlag, Berlin, 2004).

\bibitem{ol28-1302}
V.~R. Almeida, R.~R. Panepucci, and M.~Lipson, \enquote{Nanotaper for compact
  mode conversion,} Opt. Lett. \textbf{28}, 1302--1304 (2003).

\bibitem{ieee-ofc2003-v1-p249}
I.~Day, I.~Evans, A.~Knights, F.~Hopper, S.~Roberts, J.~Johnston, S.~Day,
  J.~Luff, H.~Tsang, and M.~Asghari, \enquote{Tapered silicon waveguides for
  low insertion loss highly-efficient high-speed electronic variable optical
  attenuators,} in \emph{IEEE OFC 2003}, vol.~1, pp. 249--251 (IEEE, 2003).

\bibitem{oe11-3555}
A.~Sure, T.~Dillon, J.~Murakowski, C.~Lin, D.~Pustai, and D.~Prather,
  \enquote{Fabrication and characterization of three-dimensional silicon
  tapers,} Opt. Express \textbf{11}, 3555--3561 (2003).

\bibitem{optcommun113-133}
M.~L. Gorodetsky and V.~S. Ilchenko, \enquote{High-$Q$ optical
  whispering-gallery microresonators: precession approach for spherical mode
  analysis and emission patterns with prism couplers,} Opt. Commun.
  \textbf{113}, 133--143 (1994).

\bibitem{josaA17-802}
H.~Ishikawa, H.~Tamaru, and K.~Miyano, \enquote{Microsphere resonators strongly
  coupled to a plane dielectric substrate: coupling via the optical near
  field,} J. Opt. Soc. Am. A \textbf{17}(4), 802--813 (2000).

\bibitem{ieee-jlt16-1228}
J.~Leuthold, J.~Eckner, E.~Gamper, P.~A. Besse, and H.~Melchior,
  \enquote{Multimode interference couplers for the conversion and combining of
  zero- and first-order modes,} IEEE J. Lightwave Technol. \textbf{16}(7),
  1228--1239 (1998).

\bibitem{ieee-jlt16-1680}
M.~M. Sp{\"u}hler, B.~J. Offrein, G.-L. Bona, R.~Germann, I.~Massarek, and
  D.~Erni, \enquote{A very short planar silica spot-size converter using a
  nonperiodic segmented waveguide,} IEEE J. Lightwave Technol. \textbf{16}(9),
  1680--1685 (1998).

\bibitem{ol24-723}
V.~S. Ilchenko, X.~S. Yao, and L.~Maleki, \enquote{Pigtailing the high-$Q$
  cavity: a simple fiber coupler for optical whispering-gallery modes,} Opt.
  Lett. \textbf{24}, 723--725 (1999).

\bibitem{ol20-813}
N.~Dubreuil, J.~C. Knight, D.~K. Leventhal, V.~Sandoghdar, J.~Hare, and
  V.~Lef{\`e}vre, \enquote{Eroded monomode optical fiber for whispering-gallery
  mode excitation in fused-silica microspheres,} Opt. Lett. \textbf{20},
  813--815 (1995).


\bibitem{ol22-1129}
J.~C. Knight, G.~Cheung, F.~Jacques, and T.~A. Birks, \enquote{Phase-matched
  excitation of whispering-gallery-mode resonances by a fiber taper,} Opt.
  Lett. \textbf{22}, 1129--1131 (1997).

\bibitem{ieee-ptl11-686}
M.~Cai, G.~Hunziker, and K.~Vahala, \enquote{Fiber-optic add-drop device based
  on a silica microsphere-whispering gallery mode system,} IEEE Photon.
  Technol. Lett. \textbf{11}(6), 686--687 (1999).

\bibitem{ol25-260}
M.~Cai and K.~Vahala, \enquote{Highly efficient optical power transfer to
  whispering-gallery modes by use of a symmetrical dual-coupling
  configuration,} Opt. Lett. \textbf{25}, 260--262 (2000).

\bibitem{prl91-043902}
S.~M. Spillane, T.~J. Kippenberg, O.~J. Painter, and K.~J. Vahala,
  \enquote{Ideality in a fiber-taper-coupled microresonator system for
  application to cavity quantum electrodynamics,} Phys. Rev. Lett. \textbf{91},
  043,902 (2003).

\bibitem{oe13-801}
P.~E. Barclay, K.~Srinivasan, and O.~Painter, \enquote{Nonlinear response of
  silicon photonic crystal microresonators excited via an integrated waveguide
  and fiber taper,} Opt. Express \textbf{13}, 801 (2005).

\bibitem{prb72-205318}
K.~Srinivasan, A.~Stintz, S.~Krishna, and O.~Painter,
  \enquote{Photoluminescence measurements of quantum-dot-containing
  semiconductor microdisk resonators using optical fiber taper waveguides,}
  Phys. Rev. B \textbf{72}, 205,318 (2005).

\bibitem{josab20-2274}
P.~E. Barclay, K.~Srinivasan, and O.~Painter, \enquote{Design of photonic
  crystal waveguides for evanescent coupling to optical fiber tapers and
  integration with high-$Q$ cavities,} J. Opt. Soc. Am. B \textbf{20}(11),
  2274--2284 (2003).
  

\bibitem{paddon_input_coupler_patent}
P.~J. Paddon, M.~K. Jackson, J.~F. Young, and S.~Lam, ``Photonic
  input/output port,'' U.S. Patent 7031562, Apr. 18, 2006.

\bibitem{apl77-4214}
T.~W. Ang, G.~T. Reed, A.~Vonsovici, A.~G.~R. Evans, P.~R. Routley, and M.~R.
  Josey, \enquote{Highly efficient unibond silicon-on-insulator blazed grating
  couplers,} Appl. Phys. Lett. \textbf{77}, 4214 (2000).

\bibitem{ieee-jqe38-949}
D.~Taillaert, W.~Bogaerts, P.~Bienstman, T.~F. Krauss, P.~V. Daele, I.~Moerman,
  S.~Verstuyft, K.~D. Mesel, and R.~Baets, \enquote{An out-of-plane grating
  coupler for efficient butt-coupling between compact planar waveguides and
  single-mode fibers,} IEEE J. Quantum Elect. \textbf{38}, 949--955 (2002).

\bibitem{oe14-11622}
G.~Roelkens, D.~V. Thourhout, and R.~Baets, \enquote{High efficiency
  silicon-on-insulator grating coupler based on a poly-silicon overlay,} Opt.
  Express \textbf{14}, 11,622--11,630 (2006).

\bibitem{apl87-131107}
I.-K. Hwang, S.-K. Kim, J.-K. Yang, S.-H. Kim, S.~H. Lee, and Y.-H. Lee,
  \enquote{Curved-microfiber photon coupling for photonic crystal light
  emitter,} Appl. Phys. Lett. \textbf{87}, 131,107 (2005).

\bibitem{ieee-jqe42-131}
I.-K. Hwang, G.-H. Kim, and Y.-H. Lee, \enquote{Optimization of coupling
  between photonic crystal resonator and curved microfiber,} IEEE J. Quantum
  Elect. \textbf{42}(2), 131--136 (2006).

\bibitem{oe14-1070}
C.~Grillet, C.~Smith, D.~Freeman, S.~Madden, B.~Luther-Davies, E.~C. Magi,
  D.~J. Moss, and B.~J. Eggleton, \enquote{Efficient coupling to chalcognide
  glass photonic crystal waveguides via silica optical fiber nanowires,} Opt.
  Express \textbf{14}, 1070--1078 (2006).

\bibitem{oe15-1267}
C.~Grillet, C.~Monat, C.~L.~Smith, B.~J.~Eggleton, D.~J.~Moss, 
S.~Fr{\'e}d{\'e}rick, D.~Dalacu, P.~J.~Poole, J.~Lapointe, G.~Aers, and 
R.~L.~Williams, \enquote{Nanowire coupling to photonic crystal nanocavities 
for single photon sources,} Opt. Express \textbf{15}, 1267--1276 (2007).

\bibitem{ol29-697}
P.~E. Barclay, K.~Srinivasan, M.~Borselli, and O.~Painter, \enquote{Efficient
  input and output fiber coupling to a photonic crystal waveguide,} Opt. Lett.
  \textbf{29}, 697--699 (2004).

\bibitem{oe13-1515}
M.~Borselli, T.~J. Johnson, and O.~Painter, \enquote{Beyond the Rayleigh
  scattering limit in high-Q silicon microdisks: theory and experiment,} Opt.
  Express \textbf{13}, 1515 (2005).

\bibitem{apl88-131114}
M.~Borselli, T.~J. Johnson, and O.~Painter, \enquote{Measuring the role of
  surface chemistry in silicon microphotonics,} Appl. Phys. Lett. \textbf{88},
  131,114 (2006).

\bibitem{MB_surface_passivation}
M.~Borselli, T.~J. Johnson, C.~P. Michael, M.~D.~Henry, and O.~Painter, \enquote{Surface
  encapsulation for low-loss silicon photonics,} (unpublished).

\bibitem{footnote1} 
For doublet modes, the quality factor used in Fig.~\ref{array}(c) is the average $Q$ between the two standing wave modes.

\bibitem{footnote2}
For silicon wafers with 14--20\,$\Omega\cdot$cm resistivity, free-carrier absorption~\cite{ieee-jqe23-123} limits microcavities to $Q < 9${}$\times$10$^7$--1.4$\times$10$^8$ at $\lambda\msub{o}=1532$\,nm.

\bibitem{ieee-jqe23-123}
R.~A. Soref and B.~R. Bennett, \enquote{Electrooptical effects in silicon,}
  IEEE J. Quantum Elect. \textbf{23}(1), 123--129 (1987).

\bibitem{inteltech8-129}
M.~J. Kobrinsky, B.~A. Block, J.-F. Zheng, B.~C. Barnett, E.~Mohammed,
  M.~Reshotko, F.~Roberton, S.~List, I.~Young, and K.~Cadien, \enquote{On-chip
  optical interconnects,} Intel Tech. Jour. \textbf{8}, 129--141 (2004).

\end{thebibliography}
\end{document}